\documentclass[twoside]{article}
\usepackage{psfig}
\newcommand{\beq}{\begin{equation}}
\newcommand{\eeq}{\end{equation}}

\newcommand{\forget}[1]{}
\begin{document}
\begin{titlepage}
\hfill BI-TP-97/20

\hfill June 1997

\begin{center}
{\large \bf Universality of hypercubic random surfaces} 

\vspace{1.7cm}

{\sl S. Bilke\footnotemark[1], Z. Burda\footnotemark[2] and B. Petersson\footnotemark[1]}

\vspace{0.6cm}

\footnotemark[1]
Fakult\"{a}t f\"{u}r Physik
Universit\"{a}t Bielefeld, \\
Postfach 10 01 31, Bielefeld 33501, Germany\\

\vspace{0.3cm}

\footnotemark[2]
Institute of Physics Jagellonian University,\\
ul. Reymonta 4, PL-30 059, Krak\'{o}w~16, Poland\\

\vfill
{\bf Abstract}
\end{center}
We study universality properties of the Weingarten
hyper-cubic random surfaces. Since a long time  the model of hypercubic 
random surfaces with a local restriction forbidding surface self--bendings 
was thought to be in a different universality class 
from the unrestricted model defined on the full set of surfaces.
In this paper we show that both  models
in fact belong  to the same universality class 
with the entropy exponent $\gamma=1/2$ and differ by the finite size
effects which are much more pronounced in the restricted model.
\forget{\em Hypercubic random surface models without self--bendings have been
considered 
to be in a different universality class than the unrestricted model. 
We show that this is not the case, but ub fact they have the same 
entropy exponent $\gamma = \frac{1}{2}$. This was masked, because finite size
effects are much larger. }
\noindent
\vfill

\end{titlepage}

\forget {\em In this article we study the universality properties of the
hyper-cubic random surface model. 
In particular it was indicated by numerical simulations
some time ago that the model could change universality class
by introducing constraints on the surfaces in the ensemble \cite{bb86}. 

The hyper-cubic random surface is a discrete surface built from 
square plaquettes embedded in a $Z^d$  target space. The partition function
is the sum over all possible surfaces with fixed topology,
weighted with a Boltzmann factor, where the action is proportional
to the number of plaquettes, i.e. the area of the surface.
This is the simplest geometrical candidate for the action.  
In the continuum limit it corresponds to the Nambu--Goto action
for strings.

The model was originally proposed by Weingarten \cite{dw80} as a 
non-perturbative regularization of the world sheet of bosonic strings. 
Classically it is equivalent to the Polyakov action, and the quantum 
model could therefore be similar to  two dimensional  quantum gravity 
coupled to a theory with $c=d>1$. If the model exhibits non-trivial 
behavior, it is therefore quite an attractive model  
in context with the work done to gain insight into the properties of 
gravity in the still poorly understood $c>1$ regime. }

The model of hyper-cubic surfaces was proposed by Weingarten 
\cite{dw80} as a non-perturbative regularization 
of the world sheet of bosonic strings. 
It was proven \cite{dfj84} that in general the string susceptibility
exponent $\gamma$ is equal to $1/2$ for spherical surfaces, \forget{\em
, if the surfaces are restricted
to have spherical topology,} reflecting the fact that the ensemble 
(\ref{Z}) is dominated by surfaces  having the structure of branched
polymers. It was also generally believed on grounds of the
renormalization group arguments  that one cannot change the
exponent $\gamma$ by adding local terms to the action.
Therefore it came as a surprise when  results of the numerical simulations
of a slightly modified model, where self--bendings surfaces
were excluded from the partition function, gave \forget{\em evidence for }
$\gamma = \frac{1}{4}$  \cite{bb86} in the case of a surface embedded 
in four dimensions. 

This result was subject of much  speculation \cite{a94}, because it 
provided the first non-trivial value, besides the mean field result 
$\gamma = \frac{1}{3}$, in 
the series of positive $\gamma $ discussed by Durhuus in \cite{du94}.
On the other hand the result is not strictly in conflict with the
analytic proof, as discussed in \cite{bb86}.

In this article we present new measurements of the exponent
$\gamma$.
Since the earlier measurements \cite{bb86}  \cite{ko83}
significant progress has been made
both in  computer speed and in the development 
of numerical techniques for simulations of random geometries. 
Combining this allows us to go
to systems two orders of magnitude larger than those used in the old 
measurements, and to gather very good statistics. Moreover 
we have at our disposal a very powerful method for extracting 
the exponent $\gamma$ based on the analysis of the baby-universe 
distribution.

For this work we developed a new algorithm, which
we will describe in a separate article \cite{bb97a}.
This algorithm is capable to include 
self--bendings, although for most of the data presented in this
letter they were suppressed  by adding additional code.
By reusing the $\gamma $
extraction method described in \cite{bb86} we show, that the results 
generated by our algorithm for surfaces of the size used in that
work are in agreement with the results generated by their algorithm
\cite{bbf85}. However, as we will show our
results on larger surfaces lead to completely different conclusions, namely
that the exponent $\gamma$ is consistent with $1/2$, although with
considerably stronger finite size corrections than in the case of
unrestricted surfaces.

The model is defined by the partition function~:
\beq
Z(\beta)= \sum _S \frac{1}{C(S)}e^{-\beta A } = \sum_A {\cal N}(A) e^{-\beta A}
\label{Z}
\eeq
where the first sum runs over all hypercubic surfaces without 
self--bendings and fixed (spherical) topology,  and the second runs over 
the number of plaquettes $A$, i.e. the surface area.
\forget{\em The number  $A$ denotes the number of
plaquettes, i.e. the  surface area. } The internal symmetry $C(S)$ 
is important mainly for very small surfaces.

In the large volume limit, the number of surfaces with a given area $A$ 
is expected to behave like 
\beq 
{\cal N}(A) \approx e^{\mu _{c} A} A ^{\gamma - 3} 
\label{largeN}
\eeq
Our main interest is focused on the universal entropy exponent $\gamma $,
which also characterizes the fractal structure related to baby-universes.

To extract this exponent,
the simulations are organized in a quasi--canonical way.
The area of the system fluctuates but the measurements are taken at a 
fixed value  $A_0$. We achieve this
by adding an external quadratic potential $ \gamma(A-A_0)^2 $ \cite{bbp96}
to the exponent in the partition function \ref{Z}.
The optimal value of the parameter $\beta$ is determined
experimentally in an iterative procedure which after each run estimates an
improved value of $\beta$ from the $A$ distribution.

To measure $\gamma$ we use the baby universe method
\forget{It is only possible to obtain a good estimate of $\gamma$ for large
surfaces, if one can eliminate the leading exponential term in the
measurement. For this reason, we combine the simulations with the 
baby universe method} \cite{ajt93}. This method is based 
on the observation that the distribution of baby universes 
is independent of the critical value $\mu _c$. 
By baby universe we mean a part of the surface which has 
the topology of a disc and is connected in such a way that the square of 
the disc perimeter $l$ is much smaller than the disc area $B$~: 
$l^2 \ll B$. When one looks at such a short loop  
from far enough away one sees it as a point to which
two surfaces are pinned~: one with area $B$ and the other one, the 
complementary part of the whole surface, with area $A-B$.   
For large $B$ and $A-B$ the correlations of the parts of the surface on the 
two sides of the loop are negligible and one can approximate the 
baby universe distribution by the formula \cite{jm92}~:
\beq
N_A(B) \sim {\cal N}_l(B) {\cal N}_l(A-B)
\eeq
where ${\cal N}_l(A)$ is number of discs with the loop $l$ and area $A$. 
The number ${\cal N}_l(A)$ can be expressed by the number of spherical
surfaces ${\cal N}(A)$. Take for example $l$ as the perimeter of the  
square. We have then
\beq 
{\cal N}_l(A) = (A+1){\cal N}(A+1).
\eeq
This means that one can make such a disk by removing one square from 
the lattice. Generally, for short loops one then has 
\beq
{\cal N}_l(A) \sim A {\cal N}(A)\{ 1 + o(1/A)\} .
\eeq  
In our measurements we use loops of length two and four, being, respectively,
the length of a double link and a square, {\em i.e.} the smallest possible
objects.

Using the approximate formula (\ref{largeN}) we get 
\beq
N_A(B) \sim (B(A-B))^{\gamma-2}
\label{mindis}
\eeq
for the distribution of minbus with area $B$ in a canonical ensemble with 
a total area $A$. The exponential factor $e^{\beta A}$ becomes a  
normalization factor independent of the minbu area $B$.  Taking the logarithm
of both sides and defining
\beq
\begin{array}{rcl}
y & = & \ln \left ( B(A-B) \right )^2 N_A(B) \\
x & = & \ln B(A - B)
\end{array}
\label{plot}
\eeq
one can extract $\gamma $ by fitting to the  linear equation
$y = \gamma x + b$.

\begin{figure}
\hfill
\psfig{file=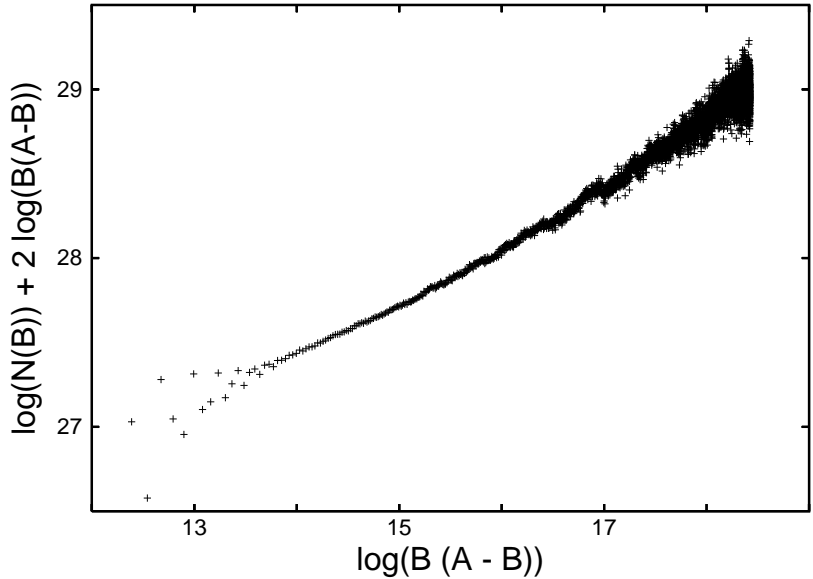,angle=270,height=11cm,rheight=8cm,rwidth=11.2cm}
\hfill

\caption{The distribution of minbu-areas for $A=10000$}
\label{mindisf}
\end{figure}

In fig(\ref{mindisf}) we show a typical minbu distribution, where we
plot $y$ versus $x$. 
The total  surface area is $A = 10000$. For small values of $B$ 
the data points do not lie on a single line but wiggle around. 
This effect comes from the combination of the symmetry factor and the
surface entropy. For large values of $B$ one can see strong statistic
noise which comes from the fact that the frequency of appearance
of the large baby universes is very small. 
For example, for the surface $A=10000$ one has $7~10^{-4}$ baby universes of 
area $B=4999$ per surface. 

\begin{figure}
\hfill
\psfig{file=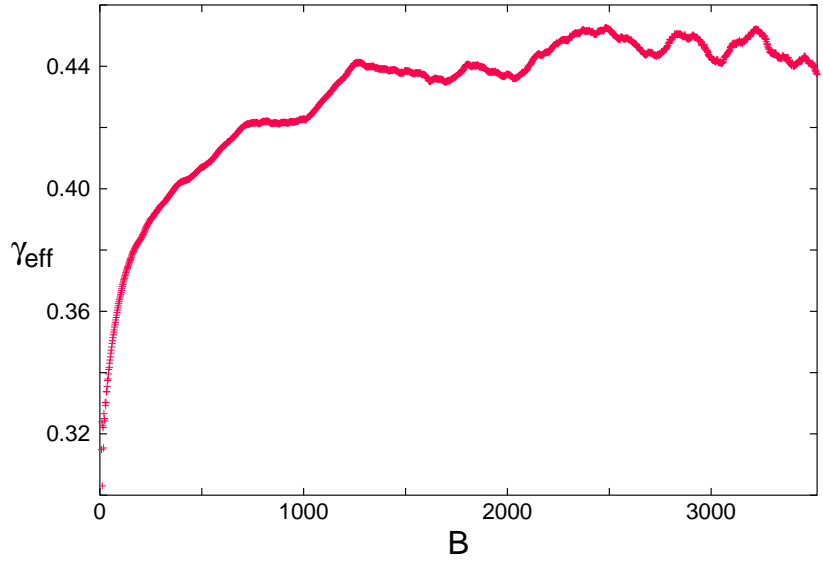,angle=270,height=11cm,rheight=8cm,rwidth=11.2cm}
\hfill

\caption{The effective  $\gamma _{eff}$ as a function of the lower cut-off.}
\label{geff}
\end{figure}

Due to finite size effects the curve  is not quite a straight line.
Therefore one actually measures an {\em effective } $\gamma _{eff}$,
which should converge to $\gamma $ with larger areas $A$.
When estimating the value of $\gamma $ 
one has to make decision which part of the distribution to
use. We extract $\gamma $
from this distribution by looking at its behavior as a function of the lower
cutoff, {\em i.e.} the smallest minbu-area used to fit $\gamma $. As the value
of $\gamma $ we take the plateau-value, as one can see for example 
in figure (\ref{geff}) for $A=10000$. It is important to note that the 
plateau value sets in for baby universes larger than $1000$. 

\begin{figure}
\hfill
\psfig{file=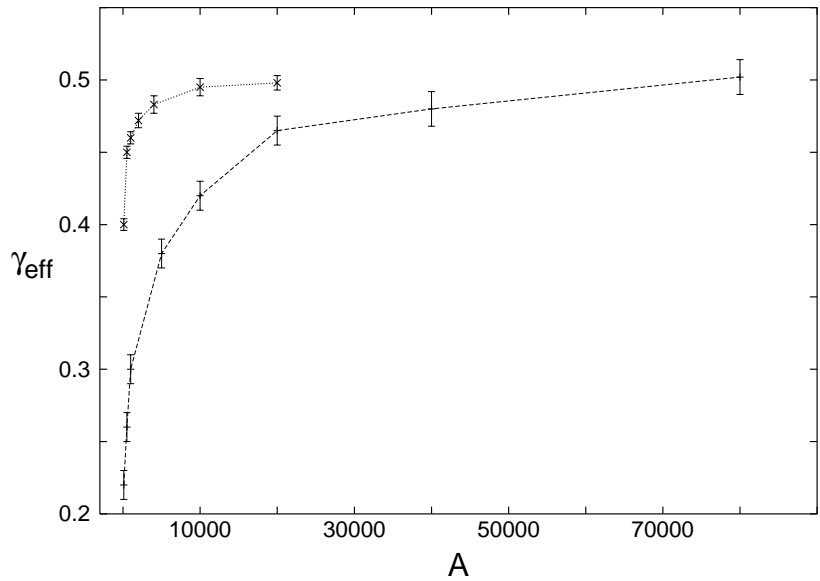,angle=270,height=11cm,rheight=8cm,rwidth=11.2cm}
\hfill

\caption{The effective entropy exponent $\gamma _{eff}$ as a function of the
canonical volume $A$ for  ensembles respectively including (upper curve) 
or excluding (lower curve) self--bendings.}
\label{gamma}
\end{figure}

We analyzed the minbu-distribution for different canonical volumes. 
In figure (\ref{gamma}) we show $\gamma _{eff}$ as a function 
of the canonical volume $A$.
For comparison we also  performed simulations for the ensemble  
containing self--bendings. As discussed above, one can for this case
show analytically that $\gamma = \frac{1}{2}$.
For the model without self--bendings (lower curve) one clearly observes stronger
finite size effects. The large area behavior of $\gamma _{eff} $, however,  is 
not affected. 

This observation is different from  the results of the 
numerical experiment reported in \cite{bb86}.
The reported value of $\gamma = 0.26$ can be found  in figure (\ref{gamma})
for volumes $A \approx 200$ which in fact correspond to the values  used in 
that work. The method used there is based on the ratio of two 
shifted histograms of the area distribution $f(A)$ for spherical
 surfaces obtained for different coupling constants in the action 
$\beta A$. As before this eliminates the unknown parameter
in the exponent which disappears as it is independent of $A$.
Thus one ends up with the formula~: 
\beq
\frac{f(A)}{f(A + \delta)} = const. \times 
  \left ( \frac{A}{A + \delta} \right )^{\gamma - 3} 
  \exp ( (\beta _1 - \beta _2)) A
\label{danaer}
\eeq
where the histograms were measured for two different
sizes $A$ and $A+\delta$.
The method is is similar to the  baby universe method  in the sense that 
one gets rid of the inconvenient unknown exponential prefactor. Here, however,
the price one has to pay is much higher
because one has to use two independent histograms, which introduces a large
statistical error.

We have repeated measurements of $\gamma$ using this method
and following parameters~:

\centerline{
\begin{tabular}{ccc}
$\beta$ & $A$  \\ 
$1.170$  &  $[170, 280]$\\
$1.160$ & $[4890, 5000]$ \\
$1.165$ & $[4890, 5000]$ \\
$1.170$ & $[4890, 5000]$ \\
\end{tabular}
}

These are  very similar to those used in the work \cite{bb86}.
For each coupling we collect ten histograms, therefore we have 
$300$ possibilities of calculating (\ref{danaer}). 
The resulting distribution of $\gamma$ - values is is quite broad and 
turns out to react rather sensitive to a change of the 
surface--areas involved in the procedure. With the above parameters we get 
$\gamma _{eff} = 0.26(10)$, which is 
in very good agreement with the results presented in \cite{bb86}.
This value is  different from the branched polymer value $1/2$ 
because $\gamma $ is measured at small volumes, where the finite 
size effects hide the real value, as can be clearly seen
in figure (\ref{gamma}).

To summarize, we have measured the value of the string
susceptibility exponent $\gamma$ for hypercubic
surfaces embedded in $d=4$-dimensional target space for
the ensemble of spherical surfaces with and  without
self--bending hinges. In both cases the results are compatible with
the branched polymer value $\gamma = 1/2$.
This means that the local infinite coupling to the external geometry, 
which suppresses self--bendings,  is not sufficient to modify the fractal
structure of the surface. 
These results  do not rule out the possibility that one can influence the
fractal structure with a generalized local coupling to the external geometry.
Obviously with an infinite coupling to the external curvature the surface has 
to be (up to finite size effects) entirely flat, {\em i.e.}
the Hausdorff-dimension is $d_h = 2$ in this case. This value is different
from $d_{h}=4$, the value obtained for the model investigated in 
this work. Therefore the fractal structure {\em  is}
affected. It remains to find out, if a transition 
appears at finite couplings. 

We have developed a very general implementation of our 
new algorithm,  which allows to study more general actions controlling
self--bendings or other local terms in the action. Moreover,
the program can be run already in the target space dimension two.
We will present results of these simulations in a forthcoming publication.

We want to thank  Joachim Tabaczek and Gudmar Thorleifsson for discussion. 
We are grateful to the HLRZ Juelich for computer 
time on the PARAGON, where parts of the simulations
were performed. Furthermore we thank the DFG under grant PE340/3--3 for
support. ZB would like to thank the Polish Committee for Scientific Research
under the grant no. 2P03B19609.

\end{document}